\documentclass[%
aip,
sd,%
amsmath,amssymb,
reprint,%
]{revtex4-1}
\usepackage[colorlinks=true,
linkcolor=red,
urlcolor=black,
citecolor=blue]{hyperref}
\usepackage{graphicx}
\usepackage{epstopdf}
\usepackage{dcolumn}
\usepackage{bm}

\begin{document}
\preprint{AIP/123-QED}
\title{Chimera states in a multilayer network of coupled and uncoupled neurons}
	\author{Soumen Majhi}
	\affiliation{Physics and Applied Mathematics Unit, Indian Statistical Institute, 203 B. T. Road, Kolkata-700108, India}
	
	\author{Matja{\v z} Perc}
	\affiliation{Faculty of Natural Sciences and Mathematics, University of Maribor, Koro{\v s}ka cesta 160, SI-2000 Maribor, Slovenia}
	\affiliation{CAMTP -- Center for Applied Mathematics and Theoretical Physics, University of Maribor, Mladinska 3, SI-2000 Maribor, Slovenia}
	
	\author{Dibakar Ghosh$^1$}\email{diba.ghosh@gmail.com}
	
	\date{\today}

	\begin{abstract}
		 We study the emergence of chimera states in a multilayer neuronal network, where one layer is composed of coupled and the other layer of uncoupled neurons. Through the multilayer structure, the layer with coupled neurons acts as the medium by means of which neurons in the uncoupled layer share information in spite of the absent physical connections among them. Neurons in the coupled layer are connected with electrical synapses, while across the two layers neurons are connected through chemical synapses. In both layers the dynamics of each neuron is described by the Hindmarsh-Rose square wave bursting dynamics. We show that the presence of two different types of connecting synapses within and between the two layers, together with the multilayer network structure, plays a key role in the emergence of between-layer synchronous chimera states and patterns of synchronous clusters. In particular, we find that these chimera states can emerge in the coupled layer regardless of the range of electrical synapses. Even in all-to-all and nearest-neighbor coupling within the coupled layer, we observe qualitatively identical between-layer chimera states. Moreover, we show that the role of information transmission delay between the two layers must not be neglected, and we obtain precise parameter bounds at which chimera states can be observed. The expansion of the chimera region and annihilation of cluster and fully coherent states in the parameter plane for increasing values of inter-layer chemical synaptic time delay are illustrated using {\it effective range} measurement. These results are discussed in the light of neuronal evolution, where the coexistence of coherent and incoherent dynamics during the developmental stage is particularly likely.		
	\end{abstract}
	
	\maketitle
	\begin{quotation}
		{\bf  Synchronization in neuronal networks is notably significant for adequate processing and transmission of information. The concurrence of synchronization and de-synchronization, popularly known as the chimera state, is associated with several neuronal functions as well. Research shows that neuronal evolution can occur in many areas of the brain simultaneously, thus affecting coupled and uncoupled layers of neurons. Since the multilayer interaction structure of the neuronal network in the brain, such as in the cortex or the thalamus, is crucial and an integral part of the system, we here consider a network consisting of two interacting layers of neurons, one with coupled and another with uncoupled neurons. We observe between-layer synchronous chimera and synchronous cluster states, depending on the coaction of both electrical and chemical synapses. We also show that inter-layer chemical synaptic delay plays a crucial role, which we elaborate in detail for all the examined asymptotic states.}
		
	\end{quotation}

\section{Introduction}
In a dynamical network of coupled oscillators, chimera states \cite{chim_rev1, chim_rev2} correspond to the exceptional concurrence of spatially separated synchronized (coherent) and desynchronized (incoherent) domains whereas the splitting of coherence into two or more domains of mutually synchronized oscillators refers to the cluster state. Initiated with Kuramoto's finding in a nonlocally coupled system of identical phase oscillators \cite{chp1}, the enthralling chimera phenomenon has been extensively studied during the past decade in a vast range of systems, e.g., in phase oscillators \cite{chp1,chp2,chp3,chp5,chp6}, neuronal models \cite{chne1,chne2,chne4,chne5,chne6,chne7,tanmoy,chne8}, chaotic systems \cite{chco1,chco2}, and Hopf normal forms \cite{chhn1,chhn2,chhn3,chhn4}. Concerning the coupling topology, this unique state has been envisaged in globally \cite{chne5,global} coupled as well as in locally connected \cite{chne5,chne7,chlo1,tanmoy} oscillators' network. Chimera states have also been realized in networks having unconventional interactions \cite{chne4,chne6,chcom,chtv,chyao}. Besides numerical and theoretical studies \cite{chspect}, chimera patterns have been widely inspected in experimental models \cite{chexp2,chexp3,chexp6} as well. As far as the emanation of cluster synchronization in oscillatory networks is concerned, there are noteworthy contributions \cite{chexp2,cl1,cl2,cl3,cl4}, even under multi layer formalism \cite{clm1} of the underlying network.
  Importantly enough, possibility in realization of such synchronous patterns under the multilayer (multiplex) structure of the network is yet to be fully explored and moreover their emergence in network consisting of coupled and uncoupled layers would be indeed quite fascinating. Regarding this, Ghosh et al. \cite{chmult2} examined coupled identical chaotic maps with delayed interactions in multiplex networks and found intra layer and layer chimera states. Asynchronous and synchronous inter-layer chimera states are confronted using nonlocally coupled phase oscillators and Hindmarsh-Rose neuron models (having only the chemical synaptic interactions) in \cite{chmult3}.
 \par  The study of multilayer networks \cite{buldyrev, gao, helbing, gomez, mathemat, nicosia, kivela, podob, boccal, podobnik, wangevo, interlayer1, interlayer2} has become exceptionally popular in recent days from various prospects due to its enormous relevancy in several complex systems arising in almost every discipline of science. Among these, multilayer structures in neural networks are certainly reasonable, and may be present based on several structural and functional properties. Ref. \cite{sheperd} suggests that the neural networks in the cortex is a multilayer one. Various precise illustrations on how multilayer network approaches to neuroimaging data can contribute innovative insights into the brain structure and evolving functions, are provided in \cite{muldoon,betzel}. Consequently allowing links of a specified single nature between the nodes and disregarding the multiple structure of a functional network like the human brain may have startling consequences on our understanding of the fundamental properties of the network \cite{zanin}.
\par Furthermore, there is a strong connection between the existence of chimera pattern and activities of neuronal systems, e.g. in various types of brain diseases \cite{braindis1, braindis2} such as Alzheimer's disease, schizophrenia and brain tumors. Chimera states are also related with the real world phenomena of unihemispheric slow-wave sleep \cite{sleep1,sleep2} of some aquatic animals (e.g. dolphins, eared seals) and some migrated birds. At the time of slow-wave sleep in these species half part of the brain is in sleep and other half remains awake. This strongly points out that in the awake part of the cerebral hemisphere, the neurons are desynchronized whereas in the sleepy part neurons are synchronized, which signifies the chimera state. Therefore, organized studies on the observation of chimera pattern in neuronal systems deserve significant attention.
\par  As noted earlier, works on detection of chimera states in neural networks include the Refs. \cite{chne1,chne2,chne4,chne5,chne6,chne7,tanmoy,chne8, chmult3}. But most of these works contemplate with either only electrical synapses (linear diffusive) or only chemical synapses (in terms of a nonlinear function). The exceptions in this regard are \cite{chne6,chne8}, where the influence of both of these synapses have been treated. In fact, in order to appropriately model a neuronal network, one should not neglect one or the other type, because a structural neuronal network comprises of connections from both electrical and chemical synapses \cite{plosce}. Thus the neurons may not be connected with each other with the same type of synapses everywhere. On the other hand, the range over which chemical synapses naturally operate is much larger than that of electrical synapses, so it would be actually feasible to connect the neurons across the layers through chemical synapses while associating the neurons in the same  layer via electrical synapses.
 \par While we ascertain non-local fashion of interaction in one of the layers, we emphasize on the matter that this non-local mode of interaction is really crucial to analyze because it may appear through a variety of circumstances in neuro-biology. For instance, the neuronal networks underlying the beautiful shell patterns of mollusks \cite{shells, murray}, in case of the formation of ocular dominance stripes observed in layer IVc of cat and monkey visual cortex \cite{murray, swindale}, spatially localized `bumps' of neuronal activity in network \cite{laing-chow, roxin-prl, gutkin}, to name only few. 
 \par Moreover, diverse neuronal developments \cite{brain,ref1,ref2,ref3,ref4} are found not only among coupled neuron groups in the same brain area, but also among uncoupled neuron groups in the same or different cortical areas. In the nervous system, activities are present not only between the coupled neurons, but also among the uncoupled neurons. Among other studies, emergence of synchrony in uncoupled neurons is explored under neuron's membrane potential stimulation \cite{uncsync} and also based on the influence of a common noisy field \cite{noise}.  Motivated by all these, in this work we focus on a neuronal network with multilayer architecture having two layers, one comprising of coupled and the other consisting of uncoupled neurons. The coupled layer is inferred to have a non-local form of interaction (through electrical synapses) while chemical synapses have been presumed to inter connect the layers. While treating such a multilayer network, we witness two intriguing phenomena, namely, between layer synchronous chimera and between layer synchronous cluster states. In these states, groups of nodes in each layer have exactly the same dynamical behavior with the same group of replica nodes in the other layer, while the individual layers possess chimera or cluster patterns respectively. Furthermore, the influence of time delay on these states due to information transmission between the  layers has also been examined by plotting several parameter regions rigorously and using effective range measurement.

\par The remaining part of this paper is organized as fol-
lows. In Sec. II, we discuss the mathematical form of the network and explain all the parameters used. Sec. IIIA  deals with the case when no  inter-layer synaptic delay is present and synchronous chimera and cluster states emerge between the  layers with non-local interaction in the coupled layer. In Sec. IIIB, we report the impact of inter-layer synaptic delay on the observed patterns in the network. Finally, we summarize our results in Sec. IV.

\section{ Mathematical form of the network}
\par
\begin{figure}
	\centerline{
		\includegraphics[scale=0.55]{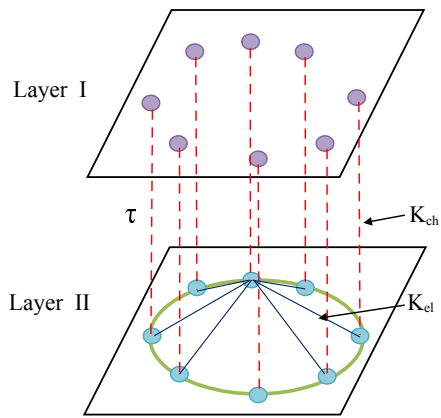}}
	\caption{  Schematic diagram of a multilayer network with two layers where upper layer (layer I) nodes are uncoupled and the lower layer (layer II) nodes are coupled nonlocally. Nonlocal interaction in the lower layer is illustrated for a single node only because of the clarity of the figure. All the other nodes  are connected to the neighboring nodes in a similar manner. Each node in the upper layer is connected to its immediate bottom node in the lower layer.}
	\label{schema}
\end{figure}

  We consider a multilayer network having two layers with same $N$ number of nodes in which the upper layer (layer I) consists of identical isolated neurons and the lower layer (layer II) is composed of non-locally coupled (in general) identical neurons. Each isolated neuron in the layer I is connected directly with one neuron (its replica) in the layer II (termed as `medium' here). The schematic diagram of the network is shown in Fig.\ref{schema}.

We assume that each node in both the  layers is having Hindmarsh-Rose neuron dynamics. The neurons in the medium are connected through electrical synapses and the isolated neurons are connected with the common medium through chemical synapses, keeping in mind the fact that the width between presynaptic and postsynaptic neurons is essentially larger at chemical synapses than electric synapses. We find the co-action of these two types of synapses playing a crucial role in the emergence of synchronous chimera and cluster patterns in the  layers.
\par Considering Hindmarsh-Rose (HR) models as the nodes of the network where both types of synapses (electrical and chemical) are present, the equations governing the dynamics of   layer I are the following:
\begin{equation}
\begin{array}{lcl}
\dot x_{i,1}=a{x_{i,1}}^2-{x_{i,1}}^3-y_{i,1}-z_{i,1}\\~~~~~~~~+K_{ch}(v_s-x_{i,1})\Gamma(x_{i,2}(t-\tau)),\\
 \dot y_{i,1}=(a+\alpha){x_{i,1}}^2-y_{i,1},\\
 \dot z_{i,1}=c(bx_{i,1}-z_{i,1}+e),
\end{array}
\end{equation}
and for the   layer II reads as
\begin{equation}
\begin{array}{lcl}
\dot x_{i,2}=a{x_{i,2}}^2-{x_{i,2}}^3-y_{i,2}-z_{i,2}\\+K_{ch}(v_s-x_{i,2})\Gamma(x_{i,1}(t-\tau))+K_{el}\sum\limits_{j={i-P}}^{i+P} (x_{j,2}- x_{i,2}),\\
\dot y_{i,2}=(a+\alpha){x_{i,2}}^2-y_{i,2},\\
\dot z_{i,2}=c(bx_{i,2}-z_{i,2}+e),
\end{array}
\end{equation}
where $(x_{i,1},y_{i,1},z_{i,1})$ and $(x_{i,2},y_{i,2},z_{i,2})$ represent the state vectors for the neurons in the  layer I and layer II respectively, $i=1, 2, . . .,N$; $N$ being the number of neurons in each of the  layers of the network and $K_{el}$, $K_{ch}$ are the coupling strengths associated to the electrical and chemical synapses respectively. As pointed earlier, the layer II is having interaction of nonlocal character. Each element of the ensemble is coupled with $P$ neighbors on both sides of a ring. The quantity $~R(=\frac{P}{N}$ or $\frac{P}{N-1})$ depending on $N$ (even or odd),  is usually termed as the coupling range. Here $\tau$ is the time-delay required to propagate the information between the  layers. This isolated HR model is a phenomenological model that exhibits all common dynamical features found in a number of biophysical modeling studies of bursting, where the variables $x_{i,k}$ represent the membrane potentials, and the variables $y_{i,k}$ and $z_{i,k}$ refer to the transport of ions across the membrane through the fast and slow channels, respectively for  layer I and II with $k=1, 2$. We assume the parameter $c$ small enough so that $z_{i,k}$ varies much slower than $x_{i,k}$ and $y_{i,k}$ ($k=1,2$). The synapses are excitatory as long as the reversal potential $v_s=2>x_{i,k}(t)$ for all $x_{i,k}(t)$ and all times $t$.
The chemical synaptic coupling function $\Gamma(x)$ is modeled by the sigmoidal nonlinear input-output function as
\begin{equation}
\begin{array}{lcl}
\Gamma(x)=\frac{1}{1+e^{-\lambda(x-\Theta_s)}},
\end{array}
\end{equation}
with $\lambda$ determining the slope of the function and $\Theta_s$ is the synaptic threshold. We choose the threshold $\Theta_s=-0.25$ so as to make every spike in the isolated neuron burst to reach the threshold and fix the value $\lambda=10$ throughout the work \cite{mthdinit}.

\section{Results}
We investigate two different cases based on  inter-layer interaction. In first case, we consider the instantaneous  inter-layer chemical synaptic coupling and later the effect of delay present in the  inter-layer chemical synaptic coupling.

\subsection{ Instantaneous inter-layer interaction}
\par
\begin{figure}
	\centerline{
		\includegraphics[scale=0.50]{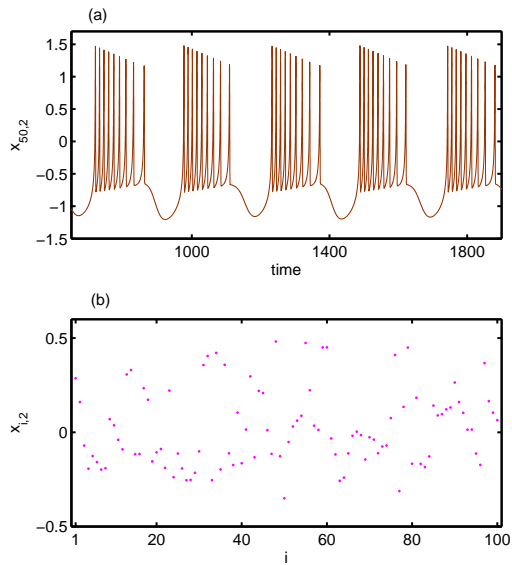}}
	\caption{  (a) Regular square-wave bursting dynamics of a particular isolated $50$-th neuron of layer II, whenever no interaction among the neurons is present (i.e. with $K_{el}=0$ and $K_{ch}=0$). (b) Typical snapshot of the membrane potentials resembling incoherent behavior of that layer with $K_{el}=0.005$, $K_{ch}=0$ and $P=30$. Here $N=100$. }
	\label{tsr}
\end{figure}

 Regular square-wave bursting dynamics of the individual neurons is observed for the choice of the set of parameter values: $a = 2.8, \alpha = 1.6, c =0.001, b = 9$, and $e = 5$, as shown in Fig.\ref{tsr}(a) (when $K_{el}=K_{ch}=0$). We hereby start with non-zero $K_{el}$ with a fixed $P$ and $K_{ch}=0$ and concentrate on a definite dynamics of  layer II  before introducing the effect of  inter-layer interaction. Whenever $K_{el}=0.005$ with $P=30$,  layer II shows incoherent dynamics, as can be noticed from a typical snapshot (Fig.\ref{tsr}(b)) of the membrane potentials in  layer II.

\begin{figure}
	\centerline{
		\includegraphics[scale=0.530]{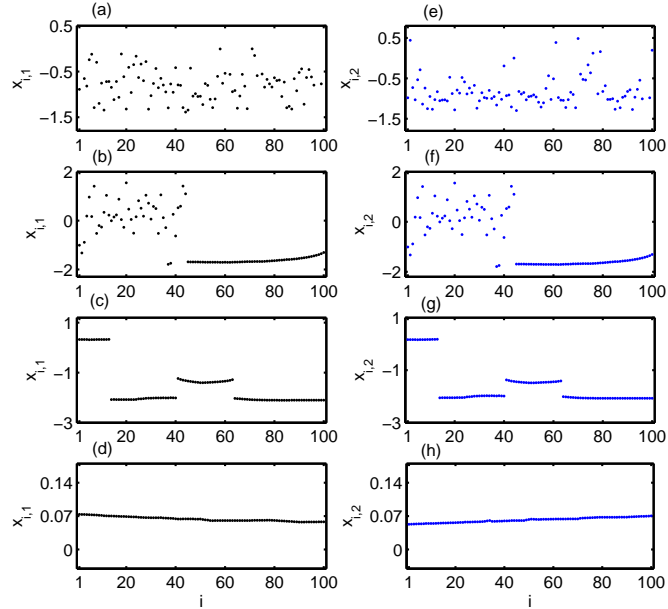}}
	\caption{  Left and right panels respectively stand for the dynamical behavior of the layer I and II depicting (a, e) incoherent state for $K_{ch}=0.5$, (b, f) chimera state for $K_{ch}=1.1$, (c, g) cluster state for $K_{ch}=2.0$ and (d, h) coherent state for $K_{ch}=3.0$. Here $K_{el}=0.005$.  }
	\label{snap}
\end{figure}

  Next, we switch the inter-layer chemical synaptic coupling strength $K_{ch}$ on, and as a consequence a transition of both the layers from disordered to coherent state via chimera and cluster state is realized. We fix the coupling radius $R=0.3$ and electric synaptic coupling strength $K_{el}=0.005$ for which the neurons in the layer II are in disordered state. At lower value of $K_{ch}=0.5,$ the neurons in both the layers are in disordered states. The snapshots of membrane potentials $x_{i,1}$ and $x_{i, 2}$ are depicted in Figs.~\ref{snap}(a) and ~\ref{snap}(e). On increasing the coupling strength to $K_{ch}=1.1$, we find the  emergence of chimera state as in Figs.~\ref{snap} (b,f). At this point, it is considerably interesting to note the appearance of synchronous chimera patterns between the layers. Here both the layers show chimera patterns with the property that the same group of replica nodes $(x_{1,1},x_{2,1},...,x_{44,1})$ and $(x_{1,2},x_{2,2},...,x_{44,2})$ form incoherent domains whereas the same group of neurons $(x_{45,1},x_{46,1},...,x_{100,1})$ and $(x_{45,2},x_{46,2},...,x_{100,2})$ exhibit coherent dynamics. On further increasing the coupling strength to $K_{ch}=2.0$,  the two  layers tend to form synchronous cluster patterns, snapshots given in Figs.~\ref{snap}(c,g). Here the layers break into four same coherent neuron groups, namely $(x_{1,k},x_{2,k},...,x_{13,k})$, $(x_{14,k},x_{15,k},...,x_{40,k})$, $(x_{41,k},x_{42,k},...,x_{63,k})$ and $(x_{64,k},x_{65,k},...,x_{100,k})$ for $k=1,2$. A bit more increment in $K_{ch}$ leads the two layers to fully coherent dynamical states, as can be seen from the snapshots in Figs.~\ref{snap} (d,h) for $K_{ch}=3.0$.

\begin{figure}
	\centerline{
		\includegraphics[scale=0.550]{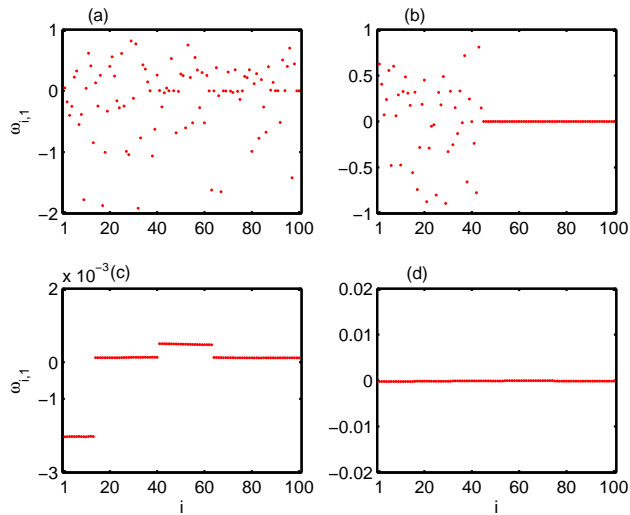}}
	\caption{  Instantaneous angular frequency of layer I neurons characterizing (a) incoherent state for $K_{ch}=0.5$, (b) chimera state for $K_{ch}=1.1$, (c) cluster state for $K_{ch}=2.0$ and (d) coherent states for $K_{ch}=3.0$. Other parameters are as in previous figures.}
	\label{af}
\end{figure}

We confirm the existence of these complex patterns by calculating instantaneous angular frequency of the $i-$th neuron of the two layers as,
$$\omega_{i, k}= \dot{\phi}_{i, k} =\frac{x_{i, k} \dot{y}_{i, k}-\dot{x}_{i, k} y_{i, k}}{x_{i, k}^2+y_{i, k}^2},$$
where $\phi_{i, k}=\mbox{arctan}(y_{i, k}/x_{i, k})$ is the geometric phase for the fast variables $x_{i, k}$ and $y_{i, k}$ of the $i$-th neuron, which is considered as a good approximation when $c$ is small ($<<1$), for $k=1,2$. The instantaneous angular frequencies reflecting incoherent, chimera, cluster and coherent states for layer I are shown in Figs.~\ref{af}(a, b, c, d) respectively. The angular frequencies corresponding to the neurons in incoherent domain are randomly scattered whereas for coherent domains that are exactly the same.  These angular frequency profiles perfectly distinguished different dynamical behaviors of the layer I. Similar frequency profiles for layer II are obtained (not shown here).

To make sure that the network behave exactly the way we claimed, we further characterize these dynamical states by using a statistical measure that uses a local standard deviation analysis, termed as strength of incoherence and defined as
\begin{equation}
\begin{array}{lcl}
 \mbox{SI}_k =1-\frac{\sum\limits_{m=1}^M s_{m,k}}{M}, ~~~~~~~~~~~~ s_{m,k}=\Theta(\delta-\sigma_k(m)),
 \end{array}
 \end{equation}
where $\sigma_k(m)$ is the local standard deviation in each of the bins (we divided the total number of neurons of both the layers into $M$ bins of equal length $n=N/M$) as follows
\begin{equation}
\begin{array}{lcl}
\sigma_k(m)=\left \langle \sqrt{\frac{1}{n} \sum\limits_{j=n(m-1)+1}^{mn} \big [\zeta_{j,k}-\left \langle \zeta_k\right\rangle \big ]^2}~~\right\rangle_t,
\end{array}
\end{equation} with $\left \langle \zeta_k \right\rangle=\frac{1}{N} \sum\limits_{i=1}^N \zeta_{i,k}(t)$;  $m=1, 2, . . .,M$ where $\zeta_{i,k}=x_{i,k}-x_{i+1,k}$, $i=1, 2, . . .,N$, and $k=1,2$. Here $\Theta(\cdot)$ is the Heaviside step function and $\delta$ is a predefined threshold.
The values $\mbox{SI}_k=1$, $\mbox{SI}_k=0$ and $0<\mbox{SI}_k<1$ correspond to the incoherent, coherent and chimera or cluster states respectively for $k$-th layer with $k=1,2$. Here we note that one is unable to distinguish between chimera and
cluster states using this $\mbox{SI}_k$ because in both  these cases, $\mbox{SI}_k$ will take similar non-zero non-unit values. At cluster state, the profile of coherent state breaks up into two or more parts and at this point the profile of $\zeta_{i, k}$ maintains smooth profile except some discontinuity at the break up points. We follow the arguments provided in \cite{SImeasure} and calculate modified strength of incoherence $\mbox{S}_k$ based on the process of removable discontinuities. The existence of disordered, chimera, cluster and coherent state are characterized by $\mbox{SI}_k$ and $\mbox{S}_k$. As mentioned above, $(\mbox{SI}_k, \mbox{S}_k)=(1, 1)$ represents a disordered state, while $(\mbox{SI}_k, \mbox{S}_k)=(0, 0)$ represents a coherent state. Further, $0<\mbox{SI}_k<1, 0<\mbox{S}_k<1$ and $0<\mbox{SI}_k<1, \mbox{S}_k=0$ represent  chimera and cluster  states respectively.
 \par
\begin{figure}
	\centerline{
		\includegraphics[scale=0.65]{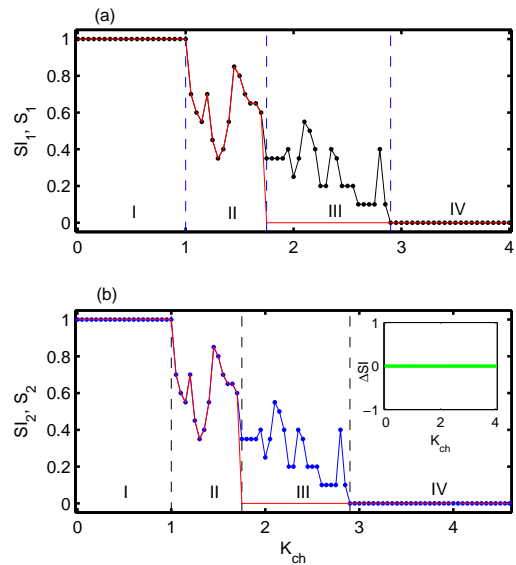}}
	\caption{  Strength of incoherence SI together with modified strength of incoherence S are plotted against inter-layer chemical synaptic coupling strength $K_{ch}$ for (a) layer I and (b) layer II.  Here $M = 20$, and  $K_{el}=0.005$.  Black line with dots and red line respectively correspond to SI and S in (a); Blue line with dots and red line respectively denote SI and S in (b). The inset of (b) depicts the variation in the SI between the layers with respect to $K_{ch}$, defined as $\Delta SI=SI_1-SI_2$.  }
	\label{SI}
\end{figure}

 Figure~\ref{SI} depicts the variation of SI of both the layers with respect to inter-layer chemical synaptic coupling strength $K_{ch}$.  We choose $ \delta = 0.05$ to calculate SI throughout the work.
 As in the Figs.~\ref{SI}(a) and ~\ref{SI}(b), in the region I=$\{K_{ch}: 0.0\le K_{ch}<1.0\}$, the values of $\mbox{SI}_k$ and $\mbox{S}_k$ $(k=1, 2)$  remain unity characterizing the incoherent (disordered) nature of the neurons in both the layers but as we increase $K_{ch}$ beyond $K_{ch}=1.0$, we observe chimera state characterized by the values $0<\mbox{SI}_k, \mbox{S}_k<1$ in the region II=$\{K_{ch}: 1.0\le K_{ch}\le 1.75\}$. With further increment in the value of $K_{ch}$, cluster states in the region III=$\{K_{ch}: 1.75\le K_{ch}\le 2.9\}$ characterized by $0<\mbox{SI}_k<1$ together with $\mbox{S}_k=0$ are realized. Finally, as $K_{ch}$ passes the value $K_{ch}=2.9$, different clusters are observed to evolve as a single one and the layers admit coherent dynamics as the values of $\mbox{SI}_k$ become zero in the region IV=$\{K_{ch}: K_{ch}> 2.9\}$. In the region II,  both the values of $\mbox{SI}_1$ and $\mbox{SI}_2$ lies in the interval (0, 1) which means both the layers are in chimera states for a wide range of $K_{ch}$. In order to identify the relation between the chimera states in the two layers, we calculate the difference between the strength of incoherences in each layer as $\Delta \mbox{SI}=\mbox{SI}_1-\mbox{SI}_2$. The non-zero values of $\Delta \mbox{SI}$ indicates asynchrony between the two layers and $\Delta \mbox{SI}=0$ represents synchronous patterns between the layers and accordingly defined as {\it synchronous chimera state} and {\it synchronous cluster state}. The inset of Fig.~\ref{SI}(b) shows the variation of $\Delta \mbox{SI}$ with $K_{ch}$.
 The zero value of $\Delta \mbox{SI}$ throughout all the values of $K_{ch}$ (In fact, the two layers maintain bounded difference between the amplitudes, i.e., the membrane potentials) together with the snapshots depicted in Fig.~\ref{snap} make the synchrony in the chimera as well as in the cluster patterns of the two layers conspicuous.
 \par
\begin{figure}
	\centerline{
	\includegraphics[scale=0.6]{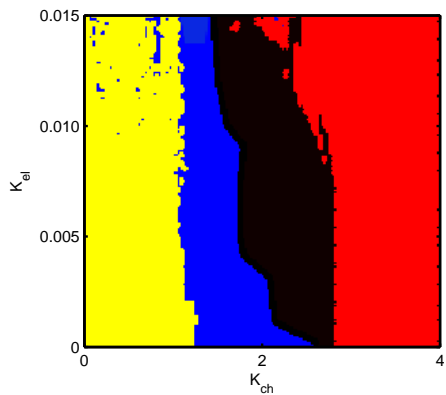}}
	\caption{ Different dynamical states in two parameters phase diagram of $K_{ch} - K_{el}$ plane where strength of incoherence and its modified form are used to distinguish different states, namely disordered, synchronous chimera, synchronous cluster and coherent states. Here $P=30$.}
	\label{kelch_p30}
\end{figure}

 In order to make the influence of both $K_{ch}$ and $K_{el}$ on the observed patterns clearer, we plot $K_{el}$ against $K_{ch}$ in Fig.~\ref{kelch_p30}. Here yellow, blue, black and red colors respectively denote incoherent, synchronous chimera, synchronous cluster and coherent states. For almost all the values of $K_{el}$ in the interval $[0.0, 0.015]$ \cite{range}, a transition from disordered state to ordered one through chimera patterns followed by cluster states can be observed. This suggest that the observation of such complex patterns as natural links between incoherence and coherence is not limited to specified values of intra layer electrical synaptic coupling strength $K_{el}$.

\par
\begin{figure}
	\centerline{
		\includegraphics[scale=0.6]{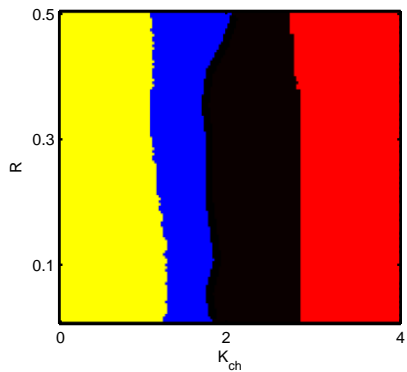}}
	\caption{ Two parameter phase diagram by simultaneously varying the inter-layer chemical synaptic coupling strength $K_{ch}$ and coupling range $R$. Here yellow, blue,  black and red colors stand for incoherent, synchronous chimera, synchronous cluster and coherent states with $K_{el}=0.005,$ $N=105$ and $M=21$.}
	\label{pkch}
\end{figure}

 To figure out the effect of the coupling range $R$ and the inter-layer interaction strength $K_{ch}$ on the dynamics of the network, we move on to the idea of formation of such intriguing patterns for every possible value of coupling radius $R$ and rigorously plot the $R-K_{ch}$ plane in Fig.~\ref{pkch}. For smaller values of $R$, the interactional framework in layer II approach to nearest neighbor coupling topology. According to the figure, synchronous chimera and cluster states arise between the layers for wide intervals of $K_{ch}$, whatever be the values of $R$. We can convert the nonlocal interaction into local and global by considering $P=1$ and $P=\frac{N-1}{2}$ for odd values of $N$. Now for $P=1,$ all the neurons in layer II are coupled with its nearest neighbor neurons only and on the other hand, for  $P=\frac{N-1}{2}$, all the neurons are globally coupled via electrical synapses. Here it is important to note that in accordance with the results present in the literature of chimera, observation of chimera state has not been possible in the sole presence of electrical synapses (linear diffusive coupling) for both of these two limiting values of $P$. However, Ref. \cite{chne5} revealed that the chimera pattern may arise for these limiting values of $P$ whenever the neurons are connected through chemical synapses only. Here, in our work, inter-layer chemical synaptic coupling function together with the underlying network structure is playing an important role for the emergence of chimera state in locally and globally coupled neurons in layer II.  Thus it is clear that not only for the intermediate values of $P$ but also for its two limiting values (corresponding to local and global coupling) the intriguing synchronous chimera with synchronous cluster states show up.

\begin{figure}
	\centerline{
		\includegraphics[scale=0.55]{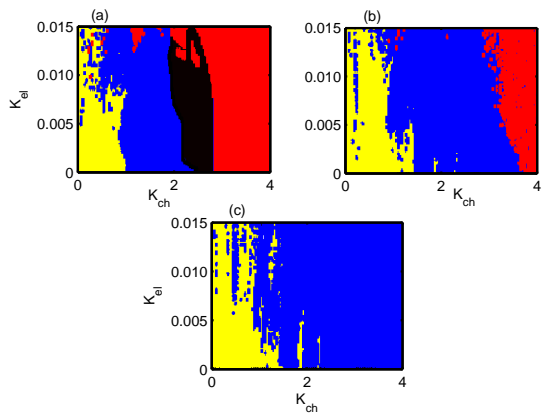}}
	\caption{  Two parameter phase diagram in $k_{ch}-k_{el}$ plane for (a) $\tau=0.5$, (b) $\tau=2.4$ and (c) $\tau=4.0$ where yellow, blue, black and red colors respectively correspond to incoherent, synchronous chimera, synchronous cluster and coherent states.}
	\label{kelch_p30_tau}
\end{figure}
\begin{figure}[b]
\centerline{
	\includegraphics[scale=0.53]{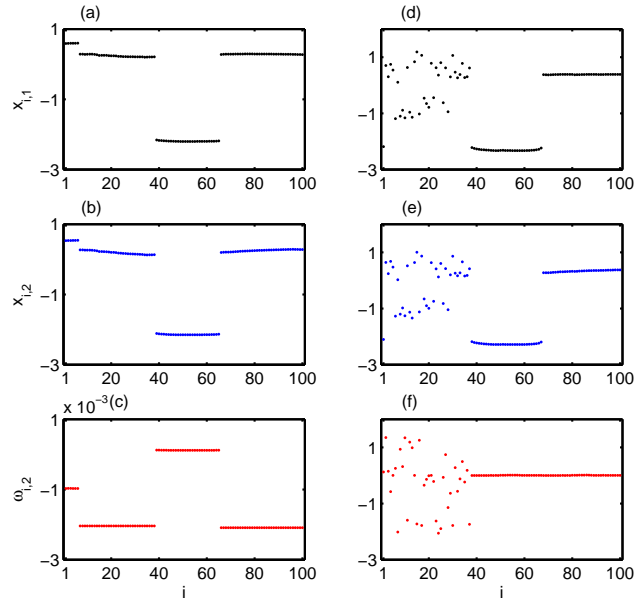}}
\caption{  Snapshots of membrane potentials resembling cluster and chimera state in (a, d) layer I and (b,e) layer II with corresponding instantaneous angular frequency $\omega_{i,2}$ of the layer II in (c) and (f) respectively. Left and right panels respectively correspond to two different values of inter-layer delay $\tau=0$ and $\tau=2.4$. Here $K_{el}=0.005$ with $K_{ch}=2.3$ fixed.  }
\label{snapdel}
\end{figure}
\subsection{ Delayed inter-layer interaction}
In neuronal networks, the effect of delay in signal transmission between different units is certain and it may genuinely arise because of the reaction times at chemical synapses. In this connection, earlier works \cite{chne8,hybridch} explained the impacts of information transmission delay in presence of both electrical and chemical synapses. So, in this subsection we investigate the development of diverse patterns of the two-layer dynamical network in response to the presence of inter-layer information transmission delay. First we consider a smaller value of the delay $\tau=0.5$ and study the dynamics of the network for the fixed $P=30$. Here again the two layers go through an evolution from incoherent to coherent dynamics (via synchronous chimera and cluster states) due to a raise in the values of inter-layer interaction strength $K_{ch}$ and this phenomenon sustains for almost every values of  intra-layer coupling strength $K_{el}$, see Fig.~\ref{kelch_p30_tau} (a). But the fact to be noted is that the values of $K_{ch}$ for which synchronous chimera and cluster patterns take place, are now different from that of the previous instantaneous interaction case (cf. Sec. IIA, Fig.~\ref{kelch_p30}).  As the observed chimera and cluster patterns between the layers are throughout synchronous, from here on we use simply `chimera' and `cluster' to indicate `synchronous chimera' and `synchronous cluster' respectively. In fact, here the chimera region in the $K_{el}-K_{ch}$ plane gets significantly enlarged whereas the cluster region in the plane gets compressed due to the introduction of the delay in the information transmission between the layers.
\par But notably, when we add more delay in the transmission, for instance with $\tau=2.4$, the cluster states in both the layers die out and chimera states come out in its arena in the parameter plane. Even some locations of the plane reflecting coherent patterns (in absence of delay), turns into spaces demonstrating chimera states now, as in Fig.~\ref{kelch_p30_tau} (b).   Finally, we boost the value of the delay further to $\tau=4.0$ while plotting the Fig.~\ref{kelch_p30_tau} (c) keeping the same bounds of $K_{el}$ and $K_{ch}$ as in the earlier cases. In this case, for the prescribed ranges of these two coupling parameters, no coherent dynamics is identified. Beside this, a much larger chunk of the plane is now illustrating chimera states compared to all the past cases.

\par The $K_{el}-K_{ch}$ parameter plane plots in Fig.~\ref{kelch_p30_tau} establishes that sufficient inter-layer delay can bring chimera in the two layers even for the values of $K_{ch}$ that was unable to do so for the instantaneous inter-layer interaction case. So, we plot some exemplary snapshots of the neurons' membrane potentials together with the angular frequencies of layer II neurons this time,  without and with time delay $\tau$ in Fig.~\ref{snapdel}. We fix $P=30$, $K_{el}=0.005$ with $K_{ch}=2.3$ for which cluster states were observed in the network for $\tau=0$ (cf. Fig.~\ref{SI}(b)). Left and right panels in Fig.~\ref{snapdel} depict the snapshots for $x_{i,1}$, $x_{i,2}$ and $\omega_{i,2}$ for the delay $\tau=0$ and $\tau=2.4$, from which a transformation from cluster pattern to chimera state by virtue of the ushering of inter-layer time delay is discernible.
\begin{figure}
	\centerline{
		\includegraphics[scale=0.7]{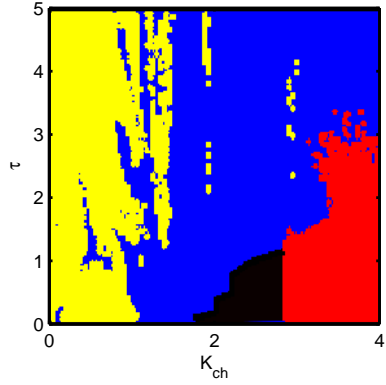}}
	\caption{  Inter-layer chemical synaptic delay $\tau$ versus the inter-layer chemical synaptic coupling strength $K_{ch}$ where yellow, blue and black and red colors respectively correspond to incoherent, synchronous chimera, synchronous cluster and coherent states for $P=30$ and $K_{el}=0.005$.}
	\label{tkch_p30_tau}
\end{figure}
\begin{figure}
	\centerline{
		\includegraphics[scale=0.45]{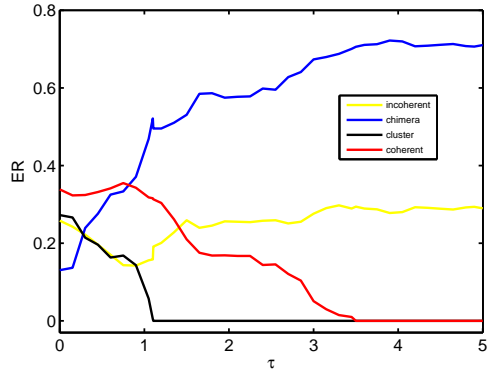}}
	\caption{  Variation in effective range (ER) of incoherent, synchronous chimera, synchronous cluster and coherent states with respect to inter-layer chemical synaptic delay $\tau$. Here $P=30$ and $T=10000$. }
	\label{er}
\end{figure}

\par While plotting Fig.~\ref{kelch_p30_tau}, we only considered some particular values of the delay $\tau$. But we now concentrate on understanding how the inter-layer delay can affect the dynamics of the network in a broader appearance and deliberately plot the $\tau-K_{ch}$ plane (with $\tau \in [0,5]$ and $K_{ch} \in [0,4]$) while keeping $P=30$ and $K_{el}=0.005$ fixed, in Fig.~\ref{tkch_p30_tau}. Here the black color region standing for cluster states come up in a very small section of the plane, in fact existing only for $\tau$ lying in $[0,1.1]$. Consequently, as $\tau$ increases, the province of $K_{ch}$ for which chimera arises (in color blue), remarkably develops. In fact, an increase in the value of $\tau$ may cause the cluster patterns and the coherent states to get transformed into chimera patterns in the parameter plane. Similar to the two-parameter phase diagrams discussed above, here also incoherent and coherent regions are shown in yellow and red colors respectively.
\begin{figure}[b]
	\centerline{
		\includegraphics[scale=0.6]{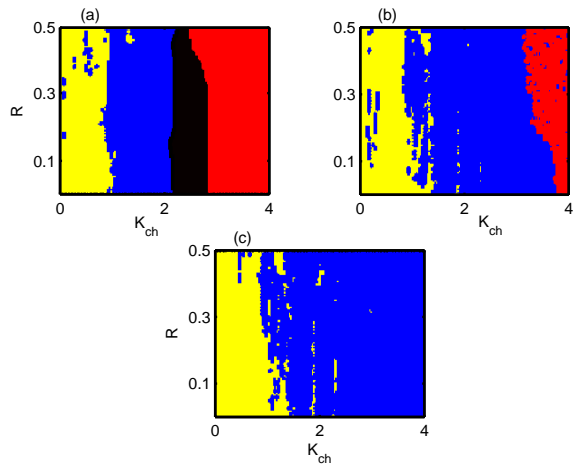}}
	\caption{ Simultaneous variation in coupling range $R$ and the inter-layer chemical synaptic coupling strength $K_{ch}$ representing incoherent, synchronous chimera, synchronous cluster and coherent regions in yellow, blue and black and red colors respectively  for $K_{el}=0.005$ with (a) $\tau=0.5$, (b) $\tau=2.4$ and (c) $\tau=4.0$. Here $N=105$.}
	\label{pkch_dly}
\end{figure}

\par For better perception about the impact of delay $\tau$ in inducing chimera pattern and eradicating cluster and coherent states in the $K_{el} - K_{ch}$ parameter plane, we calculate the effective range of all these states with increasing values of $\tau$. We integrate the system equations (1) and (2) for $T$ number of parameter points from the range $\{(K_{el},K_{ch}): K_{el}\in [0,0.015], K_{ch}\in [0,4.0]\}$ and count the number $E$ of points that reach one of these states, e.g., incoherent, chimera, cluster and coherent states. The effective range (ER) of these states are then defined to be $\frac{E}{T}$. The value of ER lies between 0 and 1. The ER=0 means there are no points in the prescribed range of parameters in $K_{el} - K_{ch}$ plane for which that particular state emerges. Basically ER represents a sort of probability of finding a particular state in the prescribed parameter plane. Figure~\ref{er} shows variation of ER of all the states against the time delay $\tau$ in which yellow, blue, black and red color curves respectively correspond to incoherent, chimera, cluster and coherent states.  It is clearly observed in this plot that after a certain threshold of $\tau=1.10$, there are no points in the parameter range for which cluster state appears. Similarly for $\tau \ge 3.5,$ the coherent states are diminished for all the values of $K_{el}$ and $K_{ch}$ in the prescribed region. The ER of chimera states in  $K_{el} - K_{ch}$ plane increases with the increasing values of $\tau$ which was also illustrated in Fig.~\ref{kelch_p30_tau}. Thus this figure perfectly demonstrates how the time-delay $\tau$ in the inter-layer interaction affects the different dynamical states.
\par Finally we move on to plot $R-K_{ch}$ plane in Fig.~\ref{pkch_dly} in presence of different values of the chemical synaptic delay $\tau$. For smaller value of $\tau=0.5$, Fig.~\ref{pkch_dly} (a) shows that the chimera region in the plane gets augmented due to the delay while the cluster zone becomes narrower (compared to the regions in Fig.~\ref{pkch}, for the case without delay). With higher value of $\tau=2.4$, cluster state does not exist any more, as in Fig.~\ref{pkch_dly} (b) and chimera pattern dominates in the $R-K_{ch}$ plane. As the delay $\tau$ in information transmission increases to $\tau=4.0$, region indicating coherence disappears as well and incoherent state followed by chimera pattern comes up  as $K_{ch}$ increases for any value of the coupling range $R$.

In this way the inevitable information transmission delay associated with the chemical synapses is enlarging the synchronous chimera domains while slowing the advancement of cluster states in the above defined neural network model of coupled and uncoupled neuron layers.

\section{Discussion and Conclusion}
  To summarize, we have put forward the notion of two synchronous complex patterns, like synchronous chimera and synchronous cluster states, by considering a  network of interacting layers of coupled and uncoupled neurons. In this work, we have investigated these states while treating different aspects of neural networks together, as the raised dynamical appearances have ample importance in the context of several neuronal developments. On one hand, various works have confirmed that multilayer frameworks in neural networks are quite likely. On the other hand, activities in the network may be present not only among coupled neurons but also between uncoupled neurons as well, so we specifically look on a structure of the network comprising of coupled and uncoupled layers of neurons. Coexistence of electrical and chemical synaptic interactions between neurons must not be neglected and hence these connections have been formed within and between the layers respectively as chemical synapses function in a longer range compared to electrical synapses. Bursting dynamics arising from Hindmarsh-Rose neuron models is used to cast the nodes in the network while keeping the coupling topology of the connected layer in non-local format (in general).
 \par 	Exceptionally, such a network structure comprising of hybrid synapses has been noticed to generate synchronous chimera and synchronous cluster patterns between the layers even when the neurons in the coupled layer are interacting locally and globally where emergence of chimera is not possible in the sole presence of electrical synaptic coupling. These states are characterized by plotting angular frequency profiles and also using the strength of incoherence for broad ranges of the parameters comprehensively. Further, the indispensable information transmission delay $\tau$ between the layers is incorporated in the network and appropriate delay has been found to induce chimera state even where the instantaneous interaction between the layers is unable to do so. We observe
 	that the cluster and coherent profiles in the parameter planes are completely eliminated once $\tau$ exceeds certain thresholds. Thus the delay may break the cluster and coherent patterns and remarkably broaden the range of chimera in the parameter space. The evidence of augmentation of the synchronous chimera region in the parameter space due to the addition of that delay has been appended in several ways, by plotting all possible parameter spaces and the effective range of all the states as well. On the grounds of the obtained results, the present work may be expected to provide crucial understanding and fundamentally deepen the perception of neuronal evolution where coexistence of coherent and incoherent dynamics of the neurons appears.

\begin{acknowledgments}
This research was supported by the Slovenian Research Agency (Grants J1-7009 and P5-0027). D.G. was supported by SERB-DST (Department of Science and Technology), Government of India (Project no. EMR/2016/001039).
\end{acknowledgments}


\begin{thebibliography}{25}
    
     \bibitem{chim_rev1}  M. J. Panaggio and D. M. Abrams, Nonlinearity {\bf 28}, R67 (2015).
    
    \bibitem{chim_rev2} B. K. Bera, S. Majhi, D. Ghosh and M. Perc, Eur. Phys. Lett. {\bf 118},  10001 (2017).
	
	\bibitem{chp1}  Y. Kuramoto and D. Battogtokh, Nonlinear Phenom. Complex Syst. {\bf 5}, 380 (2002).
	
	\bibitem{chp2}  D. M. Abrams and S. H. Strogatz, Phys. Rev. Lett. {\bf 93}, 174102 (2004).
	
	\bibitem{chp3}  D. M. Abrams, R. Mirollo, S. H. Strogatz, and D. A. Wiley, Phys. Rev. Lett. {\bf 101}, 084103 (2008).
	
	\bibitem{chp5}  A. E. Motter, Nat. Phys. {\bf 6}, 164 (2010).
	
	\bibitem{chp6}  E. A. Martens, C. R. Laing, and S. H. Strogatz, Phys. Rev. Lett. {\bf 104}, 044101 (2010).
	
	\bibitem{chne1}  I. Omelchenko, O. E. Omel’chenko, P. H\"{o}vel, and E. Sch\"{o}ll, Phys. Rev. Lett. {\bf 110}, 224101 (2013).
	
	\bibitem{chne2}  A. Vullings, J. Hizanidis, I. Omelchenko, and P. Hovel, New J. Phys. {\bf 16}, 123039 (2014).
	
	\bibitem{chne4}  I. Omelchenko, A. Provata, J. Hizanidis, E. Sch\"{o}ll, and P. H\"{o}vel, Phys. Rev. E {\bf 91}, 022917 (2015).
	
	\bibitem{chne5}  B. K. Bera, D. Ghosh, and M. Lakshmanan, Phys. Rev. E {\bf 93}, 012205 (2016).
	
	\bibitem{chne6}  J. Hizanidis, N. E. Kouvaris, G. Zamora-L\'{o}pez, A. D\'{\i}az-Guilera, and C. G. Antonopoulos, Sci. Rep. {\bf 6}, 19845 (2016).
	
	\bibitem{chne7}  B. K. Bera and D. Ghosh, Phys. Rev. E {\bf 93}, 052223 (2016).
	
	\bibitem{tanmoy}  B. K. Bera, D. Ghosh, and T. Banerjee, Phys. Rev. E {\bf 94}, 012215 (2016).
	
	\bibitem{chne8}  S. Majhi, M. Perc, and D. Ghosh, Sci. Rep. {\bf 6}, 39033 (2016).
	
	\bibitem{chco1}  I. Omelchenko, Y. Maistrenko, P. H\"{o}vel, and E. Sch\"{o}ll, Phys. Rev. Lett. {\bf 106}, 234102 (2011).
	
	\bibitem{chco2}  I. Omelchenko, B. Riemenschneider, P. H\"{o}vel, Y. Maistrenko, and E. Sch\"{o}ll, Phys. Rev. E {\bf 85}, 026212 (2012).
	
	\bibitem{chhn1}  A. Zakharova, M. Kapeller, and E. Sch\"{o}ll, Phys. Rev. Lett. {\bf 112}, 154101 (2014).
	
	\bibitem{chhn2}  L. Schmidt, K. Sch\"{o}nleber, K. Krischer, and V. Garcia-Morales, Chaos {\bf 24}, 013102 (2014).
	
	\bibitem{chhn3}  S. W. Haugland, L. Schmidt, and K. Krischer, Sci. Rep. {\bf 5}, 9883 (2015).
	
	\bibitem{chhn4}  L. Schmidt and K. Krischer, Phys. Rev. Lett. {\bf 114}, 034101 (2015).
	
	\bibitem{global} A. Yeldesbay, A. Pikovsky, and M. Rosenblum, \textit{Phys. Rev. Lett.} {\bf 112}, 144103 (2014); V. K. Chandrasekar, R. Gopal, A. Venkatesan, and M. Lakshmanan, \textit{Phys. Rev. E} {\bf 90}, 062913 (2014); K. Premalatha, V. K. Chandrasekar, M. Senthilvelan, and M. Lakshmanan, \textit{Phys. Rev. E} {\bf 91}, 052915 (2015).
	
	\bibitem{chlo1}  C. R. Laing, Phys. Rev. E {\bf 92}, 050904(R) (2015).
	
	
	\bibitem{chcom}  Y. Zhu, Z. Zheng, and J. Yang, Phys. Rev. E {\bf 89}, 022914 (2014).
	
	\bibitem{chtv}  A. Buscarino, M. Frasca, L. V. Gambuzza, and P. H\"{o}vel, Phys. Rev. E {\bf 91}, 022817 (2015).
	
	\bibitem{chyao}  N. Yao, Z. -G. Huang, Y. -C. Lai and Z. -G. Zheng, Sci. Rep. {\bf 3}, 3522 (2013).
	
	\bibitem{chspect} M. Wolfrum, O. E. Omelchenko, S. Yanchuk and Y. L. Maistrenko, Chaos {\bf 21}, 013112 (2011).
	
		
	\bibitem{chexp2}  M. R. Tinsley, S. Nkomo, and K. Showalter, Nat. Phys. {\bf 8}, 662 (2012).
	
	\bibitem{chexp3}  E. A. Martens, S. Thutupalli, A. Fourri\'{e}re, and O. Hallatschek, Proc. Natl. Acad. Sci. USA {\bf 110}, 10563 (2013).
	
		
	\bibitem{chexp6}  L. Larger, B. Penkovsky, and Y. Maistrenko, Phys. Rev. Lett. {\bf 111}, 054103 (2013).
	
	\bibitem{cl1} V. N. Belykh, G. V. Osipov, V. S. Petrov, J. A. K. Suykens and J. Vandewalle, Chaos {\bf 18}, 037106 (2008).
	

     \bibitem{cl2} P. N. McGraw and M. Menzinger, Phys. Rev. E {\bf 72}, 015101(R) (2005).


     \bibitem{cl3} L. M. Pecora, F. Sorrentino, A. M. Hagerstrom, T. E. Murphy and R. Roy, Nat. Commun. {\bf 5}, 4079 (2014).

     \bibitem{cl4} B. Chen, J. R. Engelbrecht and R. Mirollo, Phys. Rev. E {\bf 95}, 022207 (2017).


     \bibitem{clm1} S. Jalan and A. Singh, Eur. Phys. Lett. {\bf 113}, 30002 (2016).

     \bibitem{chmult2} S. Ghosh, A. Kumar, A. Zakharova and S. Jalan, Eur. Phys. Lett. {\bf 115}, 60005 (2016).
	
	\bibitem{chmult3} V. A. Maksimenko, V. V. Makarov, B. K. Bera, D. Ghosh, S. K. Dana, M. V. Goremyko, N. S. Frolov, A. A. Koronovskii, and A. E. Hramov, Phys. Rev. E {\bf 94}, 052205 (2016).
	
	\bibitem{buldyrev}
	S. V. Buldyrev, R. Parshani, G. Paul, H. E. Stanley and S. Havlin, Nature {\bf 464}, 1025 (2010).
	
	\bibitem{gao}
	 J. Gao, S. V. Buldyrev, H. E. Stanley, and	S. Havlin, Nat. Phys. {\bf 8}, 40 (2012).
	
	\bibitem{helbing}
	 D. Helbing, Nature {\bf 497}, 51 (2013).
	
	\bibitem{gomez}
	S. G\'{o}mez, A. D\'{\i}az-Guilera, J. G\'{o}mez-Garde\~{n}es, C. J. P\'{e}rez-Vicente, Y. Moreno, and A. Arenas, Phys. Rev. Lett. {\bf 110}, 028701 (2013).
	
	\bibitem{mathemat}
	M. De Domenico, A. Sol\'{e}-Ribalta, E. Cozzo, M. Kivel\"{a}, Y. Moreno, M. A. Porter, S. G\'{o}mez, and A. Arenas, Phys. Rev. X {\bf 3}, 041022 (2013).
	
	\bibitem{nicosia}
	V. Nicosia, G. Bianconi, V. Latora, and M. Barthelemy, Phys. Rev. Lett. {\bf 111}, 058701 (2013).
	
	\bibitem{kivela}
	M. Kivel\"{a}, A. Arenas, M. Barthelemy, J. P. Gleeson, Y. Moreno, and M. A. Porter, J. Complex Network {\bf 2}, 203 (2014).
	
	\bibitem{podob}
	B. Podobnik, A. Majdandzic, C. Curme, Z. Qiao, W. -X. Zhou, H. E. Stanley, and B. Li, Phys. Rev. E {\bf 89}, 042807 (2014).
	
	\bibitem{boccal}
	 S. Boccaletti, G. Bianconi, R. Criado, C. I. del Genio, J. G\'{o}mez-Garde\~{n}es, M. Romance, I. Sendi\~{n}a-Nadal, Z. Wang, M. Zanin, Phys. Rep. {\bf 544}, 1 (2014).
	
	\bibitem{podobnik}
	B. Podobnik, D. Horvatic, T. Lipic, M. Perc, J. M. Buld\'{u}, H. E. Stanley, J. R. Soc. Interface {\bf 12}, 20150770 (2015).
	
	\bibitem{interlayer1} R. Sevilla-Escoboza, I. Sendi\~{n}a-Nadal,   I. Leyva, R. Guti\'{e}rrez, J. M. Buld\'{u} and S. Boccaletti, Chaos {\bf 26}, 065304 (2016).
	
	\bibitem{interlayer2} L. V. Gambuzza, M. Frasca and J. G\'{o}mez-Garde\~{n}es, Eur. Phys. Lett. {\bf 110}, 20010 (2015).
	
	\bibitem{wangevo}
	Z. Wang, L. Wang, A. Szolnoki and M. Perc, Eur. Phys. J. B {\bf 88}, 124 (2015).
	
	\bibitem{sheperd} G. M. Sheperd, \emph{The Synaptic Organization of the Brain, 5th ed.} (Oxford University Press, New York, 2004).
	
	\bibitem{muldoon} S. F. Muldoon and D. S. Bassett, Phil. Sci. {\bf 83}, 710 (2016).
	
	\bibitem{betzel} R. F. Betzel and D. S. Bassett, NeuroImage (2016), http://dx.doi.org/10.1016/j.neuroimage.2016.11.006.
	
	\bibitem{zanin} M. Zanin, Physica A {\bf 430}, 184 (2015).
	
	\bibitem{braindis1}  R. Levy, W. D. Hutchison, A. M. Lozano, and  J. O. Dostrovsky, J. Neurosci. {\bf 20}, 7766 (2000).
		
	\bibitem{braindis2}  G. F. Ayala, M. Dichter, R. J. Gumnit, H. Matsumoto, and W. A. Spencer, Brain Res. {\bf 52}, 1 (1973).
		 	
	\bibitem{sleep1}  N. C. Rottenberg, C. J. Amlaner and S. L. Lima, Neurosci. Biobehav. Rev. {\bf 24}, 817 (2000).
		
	\bibitem{sleep2}  N. C. Rattenborg, Naturwissenschaften  {\bf 93}, 413 (2006).
	
	\bibitem{plosce}  L. R. Varshney, B. L. Chen, E. Paniagua, D. H. Hall, and D. B. Chklovskii, PLoS Comput. Biol. {\bf 7}, e1001066 (2011).
	
	\bibitem{shells} B. Ermentrout, J. Campbell, and G. Oster, Veliger {\bf 28}, 369 (1986).
	
	\bibitem{murray} J. D. Murray, \emph{Mathematical Biology} (Springer, New York, 1989).
	
	\bibitem{swindale} N. V. Swindale, Proc. R. Soc. London B {\bf 208}, 243 (1980).
	
	\bibitem{laing-chow} C. Laing and C. Chow, Neural Comput. {\bf 13}, 1473 (2001).
	
	\bibitem{roxin-prl} A. Roxin, N. Brunel, and D. Hansel, Phys. Rev. Lett. {\bf 94}, 238103 (2005).
	
	\bibitem{gutkin} B. S. Gutkin, C. R. Laing, C. L. Colby,  C. C. Chow, and G. B. Ermentrout, J. Comput. Neurosci. {\bf 11}, 121 (2001).
	
	\bibitem{brain} C. M. Gray, P. Koening and A. K. Engel, Nature {\bf 338}, 334 (1989).
	
	
	\bibitem{ref1}  B. Tadi\'{c}, M. Andjelkovi\'{c}, B. M. Boshkoska, and Z. Levnaji\'{c}, PLoS ONE {\bf 11}, e0166787 (2016).  
	
	\bibitem{ref2}  T. Kreuz, E. Satuvuori, M. Pofahl, and M. Mulansky, New J. Phys. {\bf 19}  043028 (2017).
	
	\bibitem{ref3}  P. Clusella, A. Politi, and M. Rosenblum, New J. Phys. {\bf 18}  093037 (2016). 
	
	\bibitem{ref4}  Z. Levnaji\'{c}, and A. Pikovsky, Phys. Rev. E {\bf 82}, 056202  (2010).
	
	\bibitem{uncsync} Y. P. Peng, J. Wang, Q. X. Miao, H. Y. Lu,   J. Biomedical Science and Engineering, {\bf 3}, 160 (2010).
	
	\bibitem{noise} A. B. Neiman and D. F. Russell, Phys. Rev. Lett. {\bf 88}, 138103 (2002).
	
	\bibitem{mthdinit} The fifth-order Runge-Kutta-Fehlberg integration scheme and modified Heun method are used to integrate the systems (1) and (2) with integration time step 0.01 for non-delayed and delayed cases respectively.
	
	\bibitem{SImeasure}  R. Gopal, V. K. Chandrasekar, A. Venkatesan, and M. Lakshmanan, Phys. Rev. E {\bf 89}, 052914 (2014).
	
	\bibitem{range}  The values of $K_{el}$ is chosen so small in order to make sure that in absence of $K_{ch}$, layer II remains in incoherent state. However, for $K_{el}$ much larger than the values taken with the layer II possessing incoherence, all the results have been well justified (not shown here).
	
	\bibitem{hybridch} C. Liu, J. Wang, H. Yu, B. Deng, X. Wei, K. Tsang, and W. Chan, Chaos {\bf 23}, 033121 (2013).
	
	
\end{thebibliography}
\end{document}